\documentclass[10pt,letterpaper]{article}
\usepackage[utf8x]{inputenc}
\usepackage[table]{xcolor}
\usepackage{amsmath}
\usepackage{amssymb}
\usepackage{cite}
\usepackage{hyperref}
\usepackage{nameref}
\usepackage{url}
\usepackage{array}
\usepackage{fullpage}
\usepackage{graphicx}
\usepackage{multirow}
\usepackage{authblk}
\usepackage{tabularx}
\usepackage{IEEEtrantools}
\usepackage{titlesec}
\usepackage[right]{lineno}

\begin{document}

\title{Concurrent generative models inform prediction error in the human auditory pathway}
\author[12]{Alejandro Tabas}
\author[12]{Katharina von Kriegstein}
\affil[1]{Department of Psychology, Technische Universit\"{a}t Dresden, Dresden, Germany}
\affil[2]{Max Planck Institute for Human Cognitive and Brain Sciences, Leipzig, Germany}
\date{}

\titleformat*{\section}{\large\bfseries}
\titleformat*{\subsection}{\bfseries}

\maketitle

\abstract{Predictive coding is the leading algorithmic framework to understand how expectations shape our experience of reality. Its main tenet is that sensory neurons encode prediction error: the residuals between a generative model of the sensory world and the actual sensory input. However, it is yet unclear how this scheme generalises to the multi-level hierarchical architecture of sensory processing. Theoretical accounts of predictive coding agree that neurons computing prediction error and the generative model exist at all levels of the processing hierarchy. However, there is not a current consensus of how predictions from independent models at different stages are integrated during the computation of prediction error. Here we investigated predictive processing with respect to two independent concurrent generative models in the auditory pathway using functional magnetic resonance imaging. We used two paradigms where human participants listened to sequences of either pure tones or FM-sweeps while we recorded BOLD responses in inferior colliculus (IC), medial geniculate body (MGB), and auditory cortex (AC). Each paradigm included the induction of two generative models: one based on local stimulus statistics; and another model based on the subjective expectations induced by task instruction. We used Bayesian model comparison to test whether neural responses in IC, MGB, and AC encoded prediction error with respect to either of the two generative models, or a combination of both. Results showed that neural populations in bilateral IC, MGB, and AC encode prediction error with respect to a combination of the two generative models, suggesting that the predictive architecture of predictive coding might be more complex than previously hypothesised.}

\newpage

\section{Introduction}

	Predictive coding \cite{Rao1999, Friston2003, Friston2005} is the leading theoretical framework for understanding how our expectations are integrated in our experience of reality \cite{Keller2018}. Its central assumption is that sensory processing is mediated by two kinds of neural elements: generative model units, which perform constant predictions about the future state of the sensory world, and prediction error units, which test these predictions against the sensory input \cite{Spratling2017, Keller2018}. When predictions are incorrect, prediction error units transmit the residuals to the generative model units triggering a model update. When predictions are correct, prediction error units remain silent, minimising the amount of neural activity elicited by the sensory input and optimising the neural code. 

	Sensory processing is organised hierarchically \cite{Epstein1993}. Low-level representations, located in nuclei of the subcortical sensory pathway and the cerebral cortex, encode stimuli according to their raw properties; high-level representation encode global holistic percepts that are ecologically meaningful \cite{Epstein1993, Ahissar2009, DiCarlo2012}. This hierarchical organisation is fully integrated in the predictive coding framework, which assumes that generative model units and prediction error units exist at each level of the processing hierarchy \cite{Rao1999, Friston2003, Spratling2017, Keller2018}. However, there is still no clear empirical basis to understand how predictions from the higher levels are used to inform prediction error at each subsequent lower level in the hierarchy. 
	
    Algorithms for predictive coding to date assume that a given level computes prediction error with respect to the predictions generated by the immediately higher level \cite{Spratling2017}. There are two ways in which predictions from other levels could affect prediction error: downstream and upstream propagation of predictions. In algorithms that are based on Rao and Ballard's classical formulation \cite{Rao1999} (see~\cite{Spratling2017} for a recent review) both kinds of propagation are hypothesised to be mediated by the prediction error units, which hold negative values when lower-level predictions are propagated upstream. In Rao and Ballard's scheme the propagation of predictions ultimately renders generative models along the hierarchy consistent with each other. However, for this mechanism to function, the parameters of each generative model must depend only on the history of the activation of the immediately lower prediction error unit. It is unclear how predictions could be propagated upstream to generative models that are informed by prior knowledge, known constraints on the stimulation, or cues from other sensory modalities.
    
    The later formulation from Friston \cite{Friston2003, Friston2005, Kiebel2008, Friston2009} proposes the additional existence of hidden states (dynamic variables that cannot be directly measured using brain imaging techniques \cite{Friston2009}) that could theoretically be used for upstream propagation of predictions. In this formulation,  generative models are informed by both, prediction error and the hidden states, allowing high-level generative models to integrate upstream-propagated predictions with information that does not directly depend on the history of the prediction error units. In exchange of this greater flexibility, these algorithms do not specify whether or how this influence is really exerted; indeed, computational implementations of these models to date have been constrained to two-level systems that did not exploit potential upstream nor downstream propagation of predictions (e.g.,~\cite{Friston2009}). 
    
	Empirical studies on predictive coding have so-far focused on whether, how, and where prediction error and predictions are encoded in the brain (see e.g.,~\cite{Walsh2020} for a recent review), rather than on the hierarchical organisation of the transmission of predictions. For instance, in the auditory modality, there is a vast corpus of literature exploring the mismatch negativity, assumed to encode prediction error in cortical areas (see e.g.,~\cite{Fitzgerald2020, Shiramatsu2021} for reviews). The encoding of prediction error has also been investigated specifically in auditory cortex (AC), both in humans (e.g.,~\cite{Bendixen2009, Wacongne2011, Blank2016, Ylinen2016, Stein2021, Heilbron2020}) and rodents (e.g.,~\cite{Nieto2016, Parras2017, Perez2021}), and in subcortical auditory stations: the auditory thalamus (medial geniculate body, MGB) and auditory midbrain (inferior colliculus, IC), also in humans \cite{Cacciaglia2015, Tabas2020, Tabas2021c} and rodents \cite{Parras2017, Malmierca2019, Lesicko2021}. Although some studies have specifically focused on the role of the auditory hierarchy on the generation of prediction error, the focus has so-far been restricted to the strength or prevalence of the prediction error signal, not in the generative models used to compute it (see~\cite{Tabas2021} for a review). 
	
	Thus, how predictions from independent concurrent generative models are combined to compute prediction error has not been considered before in neither the theoretical nor the empirical literature. In this study we investigate prediction error in a scenario with two independent concurrent generative models, one informed by the stimulus statistics (stats-informed) and one informed by explicit constraints conveyed in the task instructions (task-informed). In principle, both formulations of predictive coding, Rao's and Friston's, allow for predictions of the task-informed model to propagate downstream and influence prediction error at the lowest levels \cite{Spratling2017}. We have previously provided evidence that neural populations in IC, MGB, and AC (cytoarchitectonic fields Te1.0, Te1.1, Te1.2 and Te3 \cite{Morosan2001}) compute prediction error with respect to such generative model \cite{Tabas2020, Tabas2021c, Stein2021}. The aim of this study is to investigate whether the predictions from the stats-informed model are also tested by these neural populations, either by the integration of predictions from multiple levels or by a possible upstream propagation of predictions to the task-level model. We consider prediction error in the human IC, MGB, and AC, and two families of auditory stimuli: pure tones \cite{Tabas2020} and fast FM-sweeps \cite{Tabas2021c}.
    
    We anticipated three possible scenarios: First, that prediction error is computed only with respect to the stats-informed generative model. This possibility reflects the naive assumption that there is no transmission of predictions along the hierarchy (beyond the one-level-down transmission assumed in all formulations of predictive coding), and it is taken as a null-hypothesis. Second, that prediction error is computed only with respect to the task-informed generative model. This possibility assumes that predictions propagate only downstream, and that they overwrite the content of lower-level generative models. Third, that prediction error is computed with respect to a combination of both generative models. This possibility assumes that predictions propagate upstream and downstream or, alternatively, that predictions propagate downstream and are integrated with (but do not overwrite) the content of the low-level. 
    
    All three scenarios are compatible with Friston's formulation, which does not establish any constraints on whether and how predictions propagate along the processing hierarchy. The first two scenarios are compatible with Rao's formulation, whose mechanisms for prediction transmission are not well defined for generative models that are not informed by the local statistics of its associated prediction error unit. Although in theory the three scenarios are compatible with the neurophysiology of the auditory system, the idea that predictions can be transmitted further downstream than the immediately lower level is the one that reflects the physiology of the descending auditory pathway the best: corticofugal efferents stemming from AC target not only the immediately lower stage (the MGB) but also lower-level stages (the IC and the superior olivary complex) \cite{Lee2011, Schofield2011, Hackett2004}. This scenario also allows for a higher computational flexibility that transcends the linear hierarchical principles generally assumed in the predictive coding framework.
\section{Methods}

	\subsection{Experimental paradigm}

\begin{figure}[tbh!]
			\centering
			\includegraphics[width=0.75\textwidth]{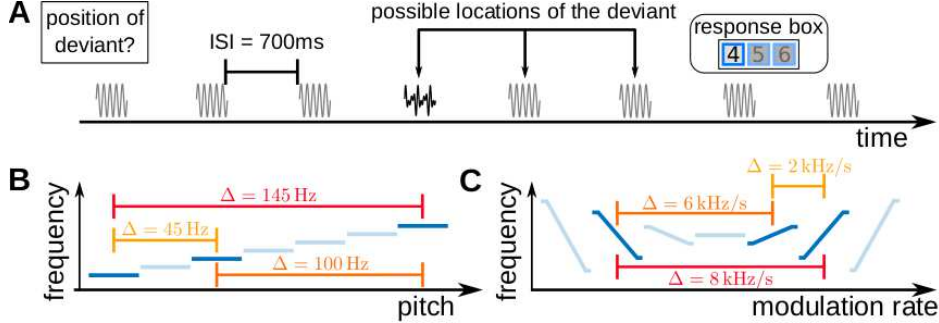}
			\vspace{1em}
			\caption{\textbf{Experimental design}. 
			A) Example of a trial. Each trial consisted of a sequence of seven repetitions of one standard (grey) and a single instance of a deviant (black). The deviant could occur in positions 4, 5, or 6 of the sequence. Participants reported, in each trial, the position of the deviant immediately after they identified it. Within a sequence, stimuli were separated by 700\,ms inter-stimulus-intervals (ISIs).  
 			B) The three pure tones used in the \emph{pure tone} experiment are displayed in dark blue. Trials were characterised by the absolute difference between the frequency of the standard and the deviant $\Delta$. 
 			C) The three FM-sweeps used in the \emph{FM-sweep} experiment are displayed in dark blue. Trials were characterised by the absolute difference between the modulation rate of the standard and the deviant $\Delta$. 
 			The stimuli schematically shown in light blue in panels B and C were not used in the experiments, and are plotted here only to contextualise the used stimuli within a family characterised by a continuously varying property (frequency in B, modulation rate in C).
			\label{fig:design}}
		\end{figure}

		A trial consisted of a sequence of eight sounds: seven repetitions of a standard and one deviant (Figure~\ref{fig:design}A). Participants were instructed to monitor the sequences and to report, as accurately and fast as possible, the position of the deviant within the sequence.

		The experimental paradigm was designed to elicit orthogonal predictions at higher- and lower-level generative models of the sensory input. We assumed that low-level predictions will be drawn by a generative model that is informed only by local stimulus statistics (i.e., the stats-informed model); namely, by neurons that monitor the identity of each sound and compute the probability of observing a given sound considering how often that tone appears in the recent stimulation history. Within a given trial, we assumed the stats-informed model to expect to find a deviant in position $n$ with a probability $P^{stats}_{n, std} = 1/8$ and a standard in position $n$ with a probability $P^{stats}_{n, std} = 1 - P_{n, std} = 1/8$. Trials were arranged in blocks of 10 that kept the same deviant and standard, to ensure that the generative models had sufficient time to infer which sound was the standard and which sound was the deviant.

		To elicit high-level predictions orthogonal to the recent stimulation history for the task-informed model, we introduced one rule: the deviant can only be located at positions 4, 5, or 6. This rule was disclosed to the participants at the beginning of the experiment, who could use it to infer task-informed predictions on the position of the deviant. The rule renders $P^{task}_{n, dev m} = 0 \forall n = [1, 2, 3, 7, 8]$, independently of the actual location of the deviant $m \in [4, 5, 6]$. The position of the deviant in each trial was pseudorandomised across the experiment so that all deviant positions were equally likely, which means that $P^{task}_{4, dev m} = 1/3 \forall m \in [4, 5, 6]$. However, if participants did not find the deviant in position 4, the deviant could only be located in positions 5 or 6; namely, $P^{task}_{5, dev m} = 1/2 \forall m \in [5, 6]$. Of the deviant is neither present in position 5, the deviant necessarily lies in  position 6, and therefore $P^{task}_{6, dev 6} = 1$. 

		The inter-trial-interval (ITI) was jittered so that deviants were separated by an average of 5 seconds, up to a maximum of 11 seconds, with a minimum ITI of 1500\,ms. This maximised the efficiency of the response estimation of the deviants \cite{Friston1999} while keeping a sufficiently long ITI to ensure that the sequences belonging to separate trials were not confounded.

		The experiment consisted in several runs of the same task. Each run contained 6 blocks of 10 trials. The 10 trials in each block used the same standard-deviant combination, so that within a block only the position of the deviant was unknown, while the identity of the deviant was known. The order of the blocks within the experiment was randomised. The position of the deviant was pseudorandomised across all trials in each run so that each deviant position happened exactly 20 times per run but an unknown amount of times per block. This constraint allowed us to keep the same prior probability for all deviant positions in each block (i.e., $P = 1/3$). In addition, there were 23 silent gaps of 5300\,ms duration (i.e., null events of the same duration as the tone sequences) randomly located in each run \cite{Friston1999}. Each run lasted around 10 minutes, depending on the reaction times of the participant.

	\subsection{Stimuli}

		All stimuli were 50\,ms long, including 5\,ms ramp-in and ramp-out Hanning windows. Stimuli were arranged in each sequence with a fixed inter-stimulus-interval of $ISI = 700$\,ms.

		There were two sets of stimuli, one based on pure tones, and one based on frequency-modulated (FM)-sweeps. Pure tones and FM-sweeps are two of the three information-bearing-elements (IBEs) \cite{Suga2012} in which meaningful acoustic signals can be linearly decomposed. We used these two sets to test whether the same principles operate across different IBE types, and thus generalise to information-bearing auditory signals.

		The pure tone set consisted of three pure tones of frequencies $f_1 = 1455\,$Hz, $f_2 = 1500\,$Hz, and $f_3 = 1600\,$Hz. With these three pure tones we built six standard-deviant combinations characterised by the absolute frequency difference between deviant and standard $\Delta = |f_{dev} - f_{std}|$. Across the experiments, participants encountered trials with three different values of $\Delta \in \{45, 100, 145\}$\,Hz (Figure~\ref{fig:design}B).

		The FM-sweep set consisted of three linear FM-sweeps, one with a descending FM (down) and two with ascending FM (up), with modulation rates $\nu_1 = -4$\,kHz/s, $\nu_2 = +2$\,kHz/s, and $\nu_1 = +4$\,kHz/s. The FM-sweeps were designed so that they elicited the same pitch percept and the same average activity across the tonotopic axis, ensuring that participants had to rely on their perception of the modulation rate to tell them apart (see~\cite{Tabas2021c} for details). Analogously to the pure tones, we used the FM-sweeps to build six standard-deviant combinations characterised by $\Delta = |\nu_{dev} - \nu_{std}| \in \{2, 4, 8\}$\,kHz/s (Figure~\ref{fig:design}B).

	\subsection{Description of the datasets}

		Data for each stimulus set was acquired with different MRI-machines and different participant cohorts. Here we describe shortly the key characteristics of each datasets; full descriptions are detailed in~\cite{Tabas2020} (pure tones) and~\cite{Tabas2021c} (FM-sweeps). Data collection of the pure tone dataset was approved by Ethics committee of the Medical Faculty of the University of Leipzig, Germany (ethics approval number 273/14-ff). Data collection of the FM-sweep dataset was approved by the Ethics committee of the Technische Universt\"{a}t Dresden, Germany (ethics approval number EK 315062019). All listeners provided written informed consent and received monetary compensation for their participation.

		Data from 19 (12 female) and 18 participants (12 female) were included in the pure tone and FM-sweeps datasets, respectively. All participants had normal hearing (thresholds equal of bellow 25\,dB in the range 250\,Hz and 8\,kHz, as measured by pure tone audiometry) and scores within the neurotypical range in screenings for developmental dyslexia and autism spectrum disorder.

    	Stimuli were presented using MATLAB (The Mathworks Inc., Natick, MA, USA; RRID:SCR\_001622) with the Psychophysics Toolbox extensions \cite{Brainard1997}. Loudness was adjusted independently for each subject before starting the data acquisition to a comfortable level. In the pure tone experiment, stimuli were delivered through an MrConfon amplifier and headphones (MrConfon GmbH, Magdeburg, Germany). In the FM-sweep experiment, stimuli were delivered through an Optoacoustics (Optoacoustics Ltd, Or Yehuda, Israel) amplifier and headphones equipped with active noise-cancellation. 

		Data from the pure tone dataset were collected using a 7-Tesla Magnetom (Siemens healthineers, Erlangen, Germany) with a spatial resolution of 1.5\,mm isotropic and temporal resolution of $TR = 1.6$\,seconds. Data from the FM-sweep dataset were collected using a 3-Tesla Trio (Siemens healthineers, Erlangen, Germany) with a spatial resolution of 1.75\,mm isotropic and temporal resolution of $TR = 1.9$\,seconds. In both cases, we used EPI sequences with partial coverage. Slices were oriented in parallel to the superior temporal gyrus such that the volumes encompassed the IC, the MGB, and the superior temporal gyrus.

		Participants from the pure tone dataset completed 4 runs in a single session (240 trials in total, 80 per deviant position). All but one participant from the FM-sweep completed 9 runs of the main experiment across three sessions (540 trials in total, 180 per deviant position); subject 18 completed only 8 runs due to technical reasons. Due to an undetected bug in the presentation code,  the first three runs of subjects 1, 2, 4, and 5; and the first six runs of subject 3 were discarded.
		
		During fMRI data acquisition, we also recorded the respiration (in the pure tone dataset) and heart rate (in both, the pure tone and FM-sweep datasets) of the participants. In addition to the functional data, we also recorded structural images of each participant using either MP2RAGE (pure tone dataset) \cite{Marques2010} or MPRAGE (FM-sweep dataset) \cite{Brant1992} protocols.
		
   		All data was preprocessed using Nipype \cite{Gorgolewski2011}, and carried out using tools of the Statistical Parametric Mapping toolbox, version 12 (SPM); Freesurfer, version 6 \cite{Fischl2002}; the FMRIB Software Library, version 5 (FSL) \cite{Jenkinson2012}); and the Advanced Normalization Tools, version 2.2.0 (ANTs) \cite{Avants2011}. All data were coregistered to the Montreal Neurological Institute (MNI) MNI152 1\,mm isotropic symmetric template. Data was first realigned and unwarped with SPM, co-registered to the participant structural image using Freesurfer, normalised to the MNI template using ANTs, and then smoothed using a 2\,mm full-width half-maximum kernel Gaussian kernel with SPM. Note that, since the resolution of the MNI space (1\,mm isotropic) was higher than the resolution of the functional data (1.5\,mm and 1.75\,mm isotropic), the transformations resulted in a spatial oversampling.

 		Physiological (respiration or/and heart rate) data was processed by the PhysIO Toolbox \cite{Kasper2017}, that computes the Fourier expansion of each component along time and adds the coefficients as covariates of no interests in the model's design matrix. All the preprocessing parameters, including the smoothing kernel size, were fixed before we started fitting the general linear model (GLM) and remained unchanged during the subsequent steps of the data analysis.

	\subsection{Regions of interest}

		We used two atlases to identify which voxels belonged to each subcortical and cortical region of interest (ROI). For the subcortical ROIs we used an \emph{in-vivo} atlas \cite{Sitek2019} that identified which voxels of the MNI space most likely cover bilateral IC and MGB. To test for potential functional specialisations of the subdivision of the MGB, we used the masks calculated in~\cite{Mihai2019} detailing the location of the ventral tonotopic axis of the nucleus. This is, to-date, the best existing approximation of the location of primary (ventral) MGB \cite{Mihai2019}. No analogous parcellation has been yet computed in the IC.

		For the cortical ROIs, we used the Morosan atlas \cite{Morosan2001}, which subdivides AC in four bilateral cortical fields using cytoarchitectural considerations. Cortical fields are identified as Te1.0, Te1.1, Te1.2, and Te3. Areas Te1.0, Te1.1, and Te1.2 are mostly located on Heschl's gyrus (Te1.1 most posterio-medial, Te1.2 most antero-lateral), and Te3 is located on the lateral surface of the superior temporal gyrus \cite{Morosan2001}. Te1.0 includes areas analogous to the core of the auditory cortex; medial Te1.0 and Te1.1 are associated to an intermediate processing stage, and Te3 is usually identified as an auditory association area \cite{Moerel2014}.

	\subsection{Bayesian model comparison}

		To evaluate whether neural responses in each of the ROIs corresponded to prediction error with respect to the stats- or task-informed generative models we used Bayesian model comparison (BMC). BMC allows to calculate the evidence for a given model of the response profile in each voxel of the region of interest. We used three BMC models that capture three different hypotheses. \emph{stats-informed}: neural responses encode prediction error with respect to a generative model that is informed by local stimulus history and statistics; \emph{task-informed}: neural responses encode prediction error with respect to the a generative model informed by the task instructions; \emph{combined}: neural responses encode prediction error with respect to a linear combination of both, statistics- and task-informed generative models. The numeric definitions of the BMC models are described below. All regressors corresponding to each of the model were normalised to have a mean of zero and variance of one across each run before fitting.

		We first computed the log-evidence for each of the three BMC models in each voxel of the ROIs per each participant using SPM via nipype. Given the model amplitude(s) ${a_n}$ and the timecourse of a voxel $y$, SPM calculates the log-evidence of the linear model $y = \beta_0 + \sum_n \beta_n a_n + \xi$, where $\beta_n$ are the linear coefficients of each regressor and $\xi$ are noise terms.

		Log-evidence maps were then combined across participants for each stimulus set using custom scripts (see~Data and code availability), following a mixed-effects-like procedure \cite{Rosa2010, Stephan2009} that results in an estimation of the log-evidence of each model for each voxel. Group-level log-evidence maps were then subtracted to compute the Bayes factor of the comparison of any two models $m_1$ and $m_2$: $K_{m_1/m_2} = e^{logEv_{m_2} - logEv_{m_1}}$.

	\subsection{Definition of the BMC models}

		\subsubsection{Modelling prediction error} 

			All BMC models assume that neural responses encode prediction error with respect to some generative model of the sensory input. We defined prediction error following~\cite{Friston2003} as the product between precision, the confidence on the prediction, and the mismatch between the expected stimulus and the actual stimulus. To model precision $\Pi$ we used the likelihood of encountering the stimulus in each position $P^{stats/task}_{n, std/dev_m}$. We assumed that the mismatch between the expected and actual stimulus would be a monotonically increasing function of the difference between the deviant and standard $\Delta$; we approximated this function to be locally linear in a neighbourhood of the set of values of $\Delta$ considering in our experiments, and to be zero if the expected and the presented stimuli were the same. As such, prediction error $\varepsilon$ is defined as:

			\begin{equation}
				\varepsilon = \sum_{s \in {stimuli}} \Pi_s \, f(\Delta, s, \text{input}), \quad
  					            f(\Delta, s, \text{input}) = \left\{\begin{array}{rl}{}
									0                 & \text{if } s = \text{input}\\
									b_0 + b_1\,\Delta & \text{if } s \neq \text{input}
								\end{array} \right. = \delta_{s, \text{input}} \, \Delta
				\label{eq:pe}
			\end{equation}

			\noindent where $\delta_{s, \text{input}}$ is the Kronecker delta, and $s \in {stimuli}$ are all the stimuli that could plausibly be heard in the next location: the standard and the deviant. For instance, if the prediction of a given generative model for a given tone is $P_{std} = 2/3, P_{dev} = 1/3$ and the tone is actually a standard, the prediction error would be $\varepsilon = P_{std} \times 0 + P_{dev} \, (b_0 + b_1 \Delta) = 1/3 \, (b_0 + b_1 \Delta)$. 

			We modelled the prediction error responses to each generative model using two regressors:

			\begin{IEEEeqnarray}{rCl}
				a^{\varepsilon}_1 & = & \sum_s \Pi_s \delta_{s, \text{input}} \label{eq:a1} \\
				a^{\varepsilon}_2 & = & \sum_s \Pi_s \delta_{s, \text{input}} \Delta_s \label{eq:a2}
			\end{IEEEeqnarray}

			\noindent Note that using these two regressors yields the linear model $y = \beta_0 + \beta_1 \sum_s \Pi_s + \beta_2 \sum_s \Pi_s \Delta_s \propto \beta_0 + \varepsilon$. 

			Regressors in Equations~\ref{eq:a1} and~\ref{eq:a2} can capture the prediction error to stimuli in positions 2-8; however, they cannot capture the responses to the first standard in each sequence. The first standard elicits prediction error with respect to the task-informed model of the sensory input not because its identity is unknown, but because its onset time is unknown. It also elicits prediction error with respect to the stats-informed model of the sensory world because it interrupts the silence that precedes it in the local stimulus history. To take into account the contributions of the first standard without tweaking the definition of $\varepsilon$ from Equation~\ref{eq:pe}, we added another regressor $a^{\varepsilon}_3 = \delta_{n, 1}$; namely, $a^{\varepsilon}_3 = 1$ for the first standard of each sequence, and $a_3 = 0$ for the remaining of the sounds in each sequence. Conversely, $a^{\varepsilon}_1$ and $a^{\varepsilon}_2$ are non-zero only for positions 2-8 within each sequence.

			While BMC models corresponding to the stats- and task-informed scenarios had 3 regressors, the BMC model that incorporates predictions of the two generative models had 5 regressors. Bayesian log-evidences penalises the addition of extra regressors, meaning that the evidence for any BMC model of a higher complexity would only be greater than the evidence for a model of lower complexity if the additional regressors explain the data better beyond what would have been expected due to overfitting of the extra free parameters.

		\subsubsection{Prediction error with respect to the stats-informed model}

\begin{figure}[tbh!]
				\centering
				\includegraphics[width=\textwidth]{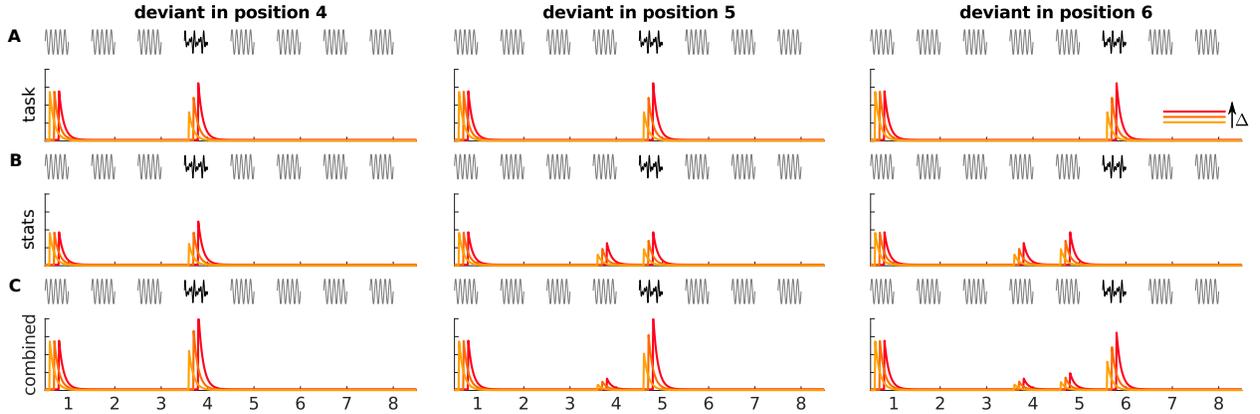}
				\vspace{1em}
				\caption{\textbf{Schematics of the models used for Bayesian model comparison}. 
				Each panel plots a possible linear combination of the regressors used in each of the four models for each of the 9 trial types (3 deviant positions $\times$ 3 values of $\Delta$) of the experiments. Plots in panel A) shows the \emph{stats} BMC model, and in panel B) the \emph{task} BMC model, and in C) the \emph{Combined} BMC model. Each coloured line corresponds to one $\Delta$ value (red corresponds to the largest delta, yellow to the lowest). The apparent delay between coloured lines is introduced only in the plots to improve visualisation. Note that the relative height of the first standard (in comparison to the deviant), and the relative weight that $\Delta$ has in the responses to the deviants are free parameters of the model.
				\label{fig:models}}
			\end{figure}

			Predictions of the stats-informed generative model were $P^{stats}_{std} = 7/8, P^{stats}_{dev} = 1/8$ for all tones in the sequence (Figure~\ref{fig:models}A). Exact values for the regressors are detailed in Table~\ref{tab:models}. 

			\begin{table}[tbh!]
			    \centering
	        	\newcommand{\D}{\Delta}

\begin{tabularx}{0.77\textwidth}{rrcccccccc}
    \multicolumn{2}{l}{\textbf{stats-informed}}    &  1  &     2     &     3     &     4     &     5     &     6     &     7     &     8     \\
    \hline 
    \multirow{3}{*}{$a_1$}    & deviant at 4 & $0$ &   $1/8$   &   $1/8$   &   $7/8$   &   $1/8$   &   $1/8$   &   $1/8$   &   $1/8$   \\
                              & deviant at 5 & $0$ &   $1/8$   &   $1/8$   &   $1/8$   &   $7/8$   &   $1/8$   &   $1/8$   &   $1/8$   \\
                              & deviant at 6 & $0$ &   $1/8$   &   $1/8$   &   $1/8$   &   $1/8$   &   $7/8$   &   $1/8$   &   $1/8$   \\
    \hline 
    \multirow{3}{*}{$a_2$}    & deviant at 4 & $0$ & $1/8\,\D$ & $1/8\,\D$ & $7/8\,\D$ & $1/8\,\D$ & $1/8\,\D$ & $1/8\,\D$ & $1/8\,\D$ \\
                              & deviant at 5 & $0$ & $1/8\,\D$ & $1/8\,\D$ & $1/8\,\D$ & $7/8\,\D$ & $1/8\,\D$ & $1/8\,\D$ & $1/8\,\D$ \\
                              & deviant at 6 & $0$ & $1/8\,\D$ & $1/8\,\D$ & $1/8\,\D$ & $1/8\,\D$ & $7/8\,\D$ & $1/8\,\D$ & $1/8\,\D$ \\
    \hline 
    \multirow{1}{*}{$a_3$}    & all deviants & $1$ &    $0$    &    $0$    &    $0$    &    $0$    &    $0$    &    $0$    &    $0$    \\
    \hline
    \hline 
    \\
    \multicolumn{2}{l}{\textbf{task-informed}}     &  1  &     2     &     3     &     4     &     5     &     6     &     7     &     8     \\
    \hline 
    \multirow{3}{*}{$a_1$}    & deviant at 4 & $0$ &    $0$    &    $0$    &   $2/3$   &    $0$    &    $0$    &    $0$    &    $0$    \\
                              & deviant at 5 & $0$ &    $0$    &    $0$    &   $1/3$   &   $1/2$   &    $0$    &    $0$    &    $0$    \\
                              & deviant at 6 & $0$ &    $0$    &    $0$    &   $1/3$   &   $1/2$   &    $0$    &    $0$    &    $0$    \\
    \hline 
    \multirow{3}{*}{$a_2$}    & deviant at 4 & $0$ &    $0$    &    $0$    & $2/3\,\D$ &    $0$    &    $0$    &    $0$    &    $0$    \\
                              & deviant at 5 & $0$ &    $0$    &    $0$    & $1/3\,\D$ & $1/2\,\D$ &    $0$    &    $0$    &    $0$    \\
                              & deviant at 6 & $0$ &    $0$    &    $0$    & $1/3\,\D$ & $1/2\,\D$ &    $0$    &    $0$    &    $0$    \\
    \hline 
    \multirow{1}{*}{$a_3$}    & all deviants & $1$ &    $0$    &    $0$    &    $0$    &    $0$    &    $0$    &    $0$    &    $0$    \\
    \hline
    \hline 
    \\
    \multicolumn{2}{l}{\textbf{combined}} &   1   &   2   &   3   &   4   &   5   &   6   &   7   &   8  \\
    \hline 
    \multirow{3}{*}{$a_1$}    & deviant at 4 & $0$ &   $1/8$   &   $1/8$   &   $7/8$   &   $1/8$   &   $1/8$   &   $1/8$   &   $1/8$   \\
                              & deviant at 5 & $0$ &   $1/8$   &   $1/8$   &   $1/8$   &   $7/8$   &   $1/8$   &   $1/8$   &   $1/8$   \\
                              & deviant at 6 & $0$ &   $1/8$   &   $1/8$   &   $1/8$   &   $1/8$   &   $7/8$   &   $1/8$   &   $1/8$   \\
    \hline 
    \multirow{3}{*}{$a_2$}    & deviant at 4 & $0$ & $1/8\,\D$ & $1/8\,\D$ & $7/8\,\D$ & $1/8\,\D$ & $1/8\,\D$ & $1/8\,\D$ & $1/8\,\D$ \\
                              & deviant at 5 & $0$ & $1/8\,\D$ & $1/8\,\D$ & $1/8\,\D$ & $7/8\,\D$ & $1/8\,\D$ & $1/8\,\D$ & $1/8\,\D$ \\
    \hline 
    \multirow{3}{*}{$a_3$}    & deviant at 4 & $0$ &    $0$    &    $0$    &   $2/3$   &    $0$    &    $0$    &    $0$    &    $0$    \\
                              & deviant at 5 & $0$ &    $0$    &    $0$    &   $1/3$   &   $1/2$   &    $0$    &    $0$    &    $0$    \\
                              & deviant at 6 & $0$ &    $0$    &    $0$    &   $1/3$   &   $1/2$   &    $0$    &    $0$    &    $0$    \\
    \hline 
    \multirow{3}{*}{$a_4$}    & deviant at 4 & $0$ &    $0$    &    $0$    & $2/3\,\D$ &    $0$    &    $0$    &    $0$    &    $0$    \\
                              & deviant at 5 & $0$ &    $0$    &    $0$    & $1/3\,\D$ & $1/2\,\D$ &    $0$    &    $0$    &    $0$    \\
                              & deviant at 6 & $0$ &    $0$    &    $0$    & $1/3\,\D$ & $1/2\,\D$ &    $0$    &    $0$    &    $0$    \\
    \hline 
    \multirow{1}{*}{$a_5$}    & all deviants & $1$ &    $0$    &    $0$    &    $0$    &    $0$    &    $0$    &    $0$    &    $0$    \\
    \hline
    \hline 
\end{tabularx}
				\caption{\textbf{Amplitudes of the models used for Bayesian Model Comparison}. 
				Amplitudes of the linear models used for BMC. For each of the three models, we computed the log-evidence that each the responses in each voxel $y \sim \sum_n \beta_n a_n$, where $\beta_n$ are the free parameters of the model.
				\emph{Stats-informed} assumes that responses encode prediction error with respect to a generative model of the sensory input that is informed by local stimulus history and statistics. \emph{Task-informed} assumes that responses encode prediction error with respect to a generative model that is informed by the task instructions. \emph{Combined} assumes that responses encode prediction error with respect to a linear combination of the predictions of the stats- and task-informed generative models of the sensory input. All regressors were normalised (mean of zero, variance of one) prior fitting.
				\label{tab:models}}
			\end{table}

		\subsubsection{Prediction error with respect to the task-informed model}

			Predictions of the task-informed generative model depended on the position of the incoming sound $n$ (Figure~\ref{fig:models}B). For positions $n \in \{2, 3, 7, 8\}$, the generative model predictions were $P^{task}_{n, std} = 1, P^{task}_{n, dev} = 0$; for position $n = 4$, they were $P^{task}_{4, dev} = 1/3, p^{task}_{4, std} = 2/3$. Predictions for positions $n \in \{5, 6\}$ depended on the actual position of the deviant $m$. If the deviant was at position $m = 4$, participants would not expect any other deviant in positions 5 and 6, and thus the generative model predictions would be $P^{task}_{n, dev} = 0, P^{task}_{n, std} = 1 \forall n \in \{5, 6\}$. If the deviant was not in position $m = 4$, the generative model predictions of $n = 5$ would be $P^{task}_{5, dev} = P^{task}_{5, std} = 1/2$. If the deviant was in position $m = 5$, predictions for $n = 6$ would be $P^{task}_{6, dev} = 0, P^{task}_{6, std} = 1$. Last, if the deviant was neither in position $m = 5$, the predictions of the generative model for $n = 6$ would be $P^{task}_{6, dev} = 1, P^{task}_{6, std} = 0$. Exact values for the regressors are detailed in Table~\ref{tab:models}. 

			With the exception of the expected responses to the first standard, the regressor $a_2$ of this BMC model is identical to the \emph{predictive coding} hypothesis in~\cite{Tabas2020, Tabas2021c, Stein2021}.

		\subsubsection{Prediction error with respect to a combination of both generative models}

			Last we consider that both, stats- and task-informed generative models could contribute to the computation of prediction error. We assumed that predictions would be a linear combination of the predictions of both models (Figure~\ref{fig:models}C); or, similarly, that the neural responses would be a linear combination of the prediction error expected by each generative model (note that the dependence of $\varepsilon$ on $\Pi$ in Equation~\ref{eq:pe} is linear). We modelled this scenario by adding the regressors $a^{\varepsilon}_1, a^{\varepsilon}_2$ corresponding to each of the two generative models. Since the responses to the first standard co-vary in both, stats- and task-informed generative model scenarios, we added only one regressor for the first standard of each sequence.

	\subsection{Measuring the temporal signal-to-noise ratio (tSNR)}

		To test whether the results were influenced by the temporal signal-to-noise ratio (tSNR) of the data, we used nipype's native \emph{confound} toolbox. We computed the tSNR in the exact same preprocessed data we used as input for the BMC analysis. Using the raw preprocessed data provided us with a fair estimation of the tSNR that was unbiased with respect to the BMC models.

	\subsection{Correlational analyses}

		All correlations reported in Results were Pearson's correlations computed across the voxels in each ROI. This means that the number of samples in each correlation is the number of voxel in the ROI. All $p$-values were Holm-Bonferroni corrected for the number of ROIs ($N=4$ in the analyses on subcortical regions, $N=10$ in the analyses of cerebral cortex). Results were deemed statistically significant when the corrected $p < 0.05$. 
	
\section{Results}

    \subsection{Concurrent generative models are combined to compute prediction error in the subcortical auditory pathway}

\begin{figure}[htbp]
            \centering
            \includegraphics[width=\textwidth]{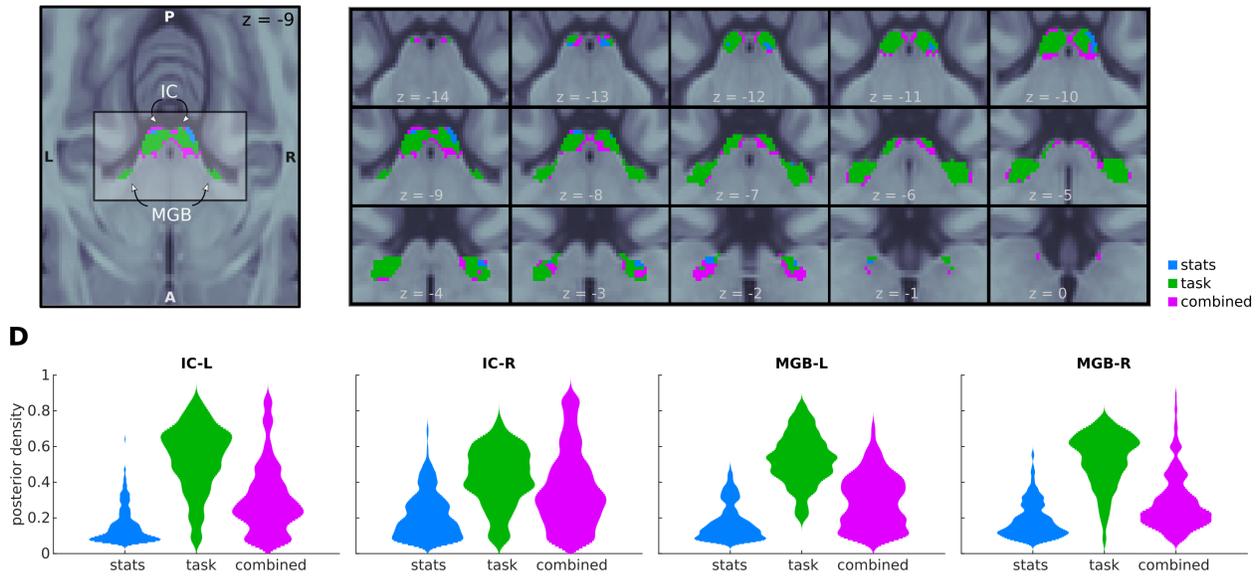}
            \caption{\textbf{Bayesian Model Comparison in IC and MGB.} A, C) Maps detailing which BMC model that best explain the responses to pure tones (A) and FM-sweeps (B) in each of the voxels of the IC and MGB ROIs. Colours indicate the BMC model with the highest posterior density at each voxel of the IC and MGB ROIs. Blue voxels are best explained by the \emph{stats-informed} BMC model, green voxels by the \emph{task-informed} BMC model, and purple voxels by the \emph{combined} BMC model. B, D) Distributions (kernel-density estimations) of the posterior densities of each model across voxels of the IC and MGB rois for the pure tone (B) and FM-sweep (D) stimuli.
    		\label{fig:bestmodels}}
        \end{figure}

    	Large sections of the IC and MGB displayed responses that were best explained by the \emph{task-informed} and the \emph{combined} BMC models, both for pure tones (Figure~\ref{fig:bestmodels}A and~B) and FM-sweeps (Figure~\ref{fig:bestmodels}C and~D). The \emph{combined} BMC model was the most prevalent explanation of the data in the four subcortical ROIs for the responses to pure tones (83\%, 76\%, 88\% and 91\% of the voxels of the left IC, right IC, left MGB, and right MGB, respectively). In the responses to FM-sweeps, populations best explained by the \emph{combined} BMC model were smaller but topologically compact (Figure~\ref{fig:bestmodels}D; 22\%, 34\%, 18\% and 18\% of the voxels of the left IC, right IC, left MGB, and right MGB, respectively). 

    	Conversely, the \emph{task-informed} BMC model was the most prevalent explanation for the responses to FM-sweeps (76\%, 56\%, 79\% and 78\% of the left IC, right IC, left MGB, and right MGB, respectively), and it explained the responses to pure tones in smaller but topologically compact populations (Figure~\ref{fig:bestmodels}B; 13\%, 19\%, 12\% and 6\% of the left IC, right IC, left MGB, and right MGB, respectively). The \emph{stats-informed} BMC model was the best explanation of the data in a few voxels scattered across the ROIs for both stimulus families with a prevalence under 5\% in almost all ROIs. The exception was the left-IC, where the stats-informed model best explained the responses to FM-sweeps in 10\% of the voxels.

    	The prevalence of task-informed and combined BMC models in different sections of the nuclei may indicate that there is not a unique strategy to propagate high-level predictions on the sensory input to low-level processing stages, but that different strategies are used in different neural populations. Another possibility is that voxels best explained by the \emph{task-informed} model are those voxels in which the BOLD responses are noisier in comparison to those in other regions. Since the \emph{combined} BMC model has two free parameters more than the \emph{task-informed} BMC model, the former needs to provide a much better explanation of the data than the latter to yield a similar log-evidence. Voxels with poorer tSNR would present higher mean-square-errors with respect to the model fits, which might have primed the presence of the \emph{task-informed} BMC model. 
    	
    	To test if that was the case we computed the correlation between the tSNR and the posterior density of the \emph{combined} BMC model across the ROIs. We found a significantly positive correlation in the three of the four ROIs for the pure tone data ($\rho \in [0.53, 0.71], p < 10^{-20}$ in IC-L, IC-R, and MGB-R; $\rho = -0.07, p = 0.2$ in MGB-L) and in all four ROIs for the FM-sweep data ($\rho \in [36, 71], p < 10^{-9}$). These results suggest that the lower tSNRs may be the reason why the \emph{combined} BMC model was not the best explanation for the data across the entire nuclei of the subcortical pathway.
    	
    	Although the general prevalence of the \emph{task-informed} and \emph{combined} models were different for the responses elicited by pure tones and FM-sweeps, they seem to follow a similar topographic organisation: populations best explained by the \emph{combined} model are located more centrally in the ICs, and more dorsally in the MBGs. To quantify if the occurrence of the \emph{task-informed} BMC model was consistent across the two stimulus families, we compared the distribution of the $K_{combined/task}$ associated to the responses to pure tones and FM-sweeps. Distribution of $K$ factors was significantly correlated in the left IC ($\rho = 0.2$, $p = 6\times10^{-4}$), right IC ($\rho = 0.31, p = 5\times10^{-8}$), and right MGB ($\rho = 0.21, p = 2.4\times10^{-3}$), but not in the left MGB ($\rho = -0.08$, $p = 0.16$). 

        However, the correlation between $K_{combined/task}$ in the pure tone and FM-sweep data could have also be driven by a similar distribution of the tSNR in both datasets. To test if that was the case, we computed the correlation between the tSNR in each dataset in the four ROIs. Despite the datasets being collected at different field strengths, with different EPI sequences and on different participant cohorts, tSNRs where highly correlated in the four ROIs ($\rho \in [0.63, 0.89], p < 10^{-32}$). These results further support the hypothesis that the \emph{combined} BMC model is only outperformed by the \emph{task-informed} BMC model in voxels for which the tSNR is not strong enough to find a suitable fit of the 5 parameters of the \emph{combined} BMC model.

    \subsection{Prediction error in the MGB is consistent across physiological subdivisions}

        The auditory pathway subdivided in primary (central section of the IC and ventral MGB) and secondary (cortex of the IC, and medial and dorsal MGB) subdivisions \cite{Hu2003}. Neurons in primary subdivisions narrowly tuned frequency responses and are responsible for the transmission of bottom-up information; neurons in secondary subdivisions present wider tuned frequency responses and are thought to be involved in multisensory integration \cite{Hu2003}. One possibility is that the functional parcellations described in Figure~\ref{fig:bestmodels} correspond to this physiological arrangement. Neural populations responding according to the \emph{combined} BMC model do indeed seem to be located towards the cortex of the ICs, although lower tSNRs are generally expected in outflanks of the nuclei.

\begin{figure}[htbp]
            \centering
            \includegraphics[width=\textwidth]{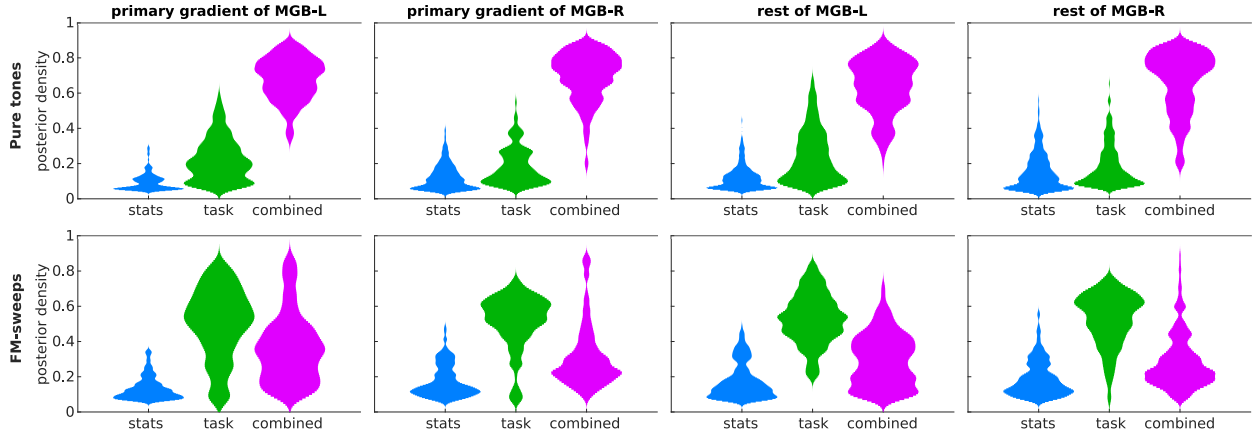}
            \caption{\textbf{Prevalence of each model in primary and secondary MGB.} Distributions (kernel-density estimations) of the posterior densities of each BMC model across voxels of the MGB subdivisions from~\cite{Mihai2019} for the pure tone and FM-sweep stimuli. Distributions are qualitatively comparable in the primary tonotopic gradient (putatively ventral MGB) and the remaining of the nuclei (putatively dorsal and medial MBG). 
            \label{fig:subdivisions}}
        \end{figure}

        Imaging subdivisions of the IC and MGB in humans is remarkably challenging \cite{Moerel2015, Mihai2019}. To-date, there is no available parcellation of the human IC into primary and secondary subdivisions; however, \cite{Mihai2019} managed to identify a ventral tonotopic gradient in the MGB that putatively corresponds to its primary subdivision. Here, we used this parcellation to assess whether neural populations in primary and secondary subdivisions of the MGB are more tuned towards the \emph{task-informed} or \emph{combined} BMC model (Figure~\ref{fig:subdivisions}). Results show that both BMC models are similarly prevalent in both subdivisions, indicating that, at least in the MGB, the functional parcellation described in Figure~\ref{fig:bestmodels} does not correspond to the physiological parcellations of the nuclei. 

    \subsection{Concurrent generative models are combined to compute prediction error in auditory cortex}

\begin{figure}[p]
            \centering
            \includegraphics[width=\textwidth]{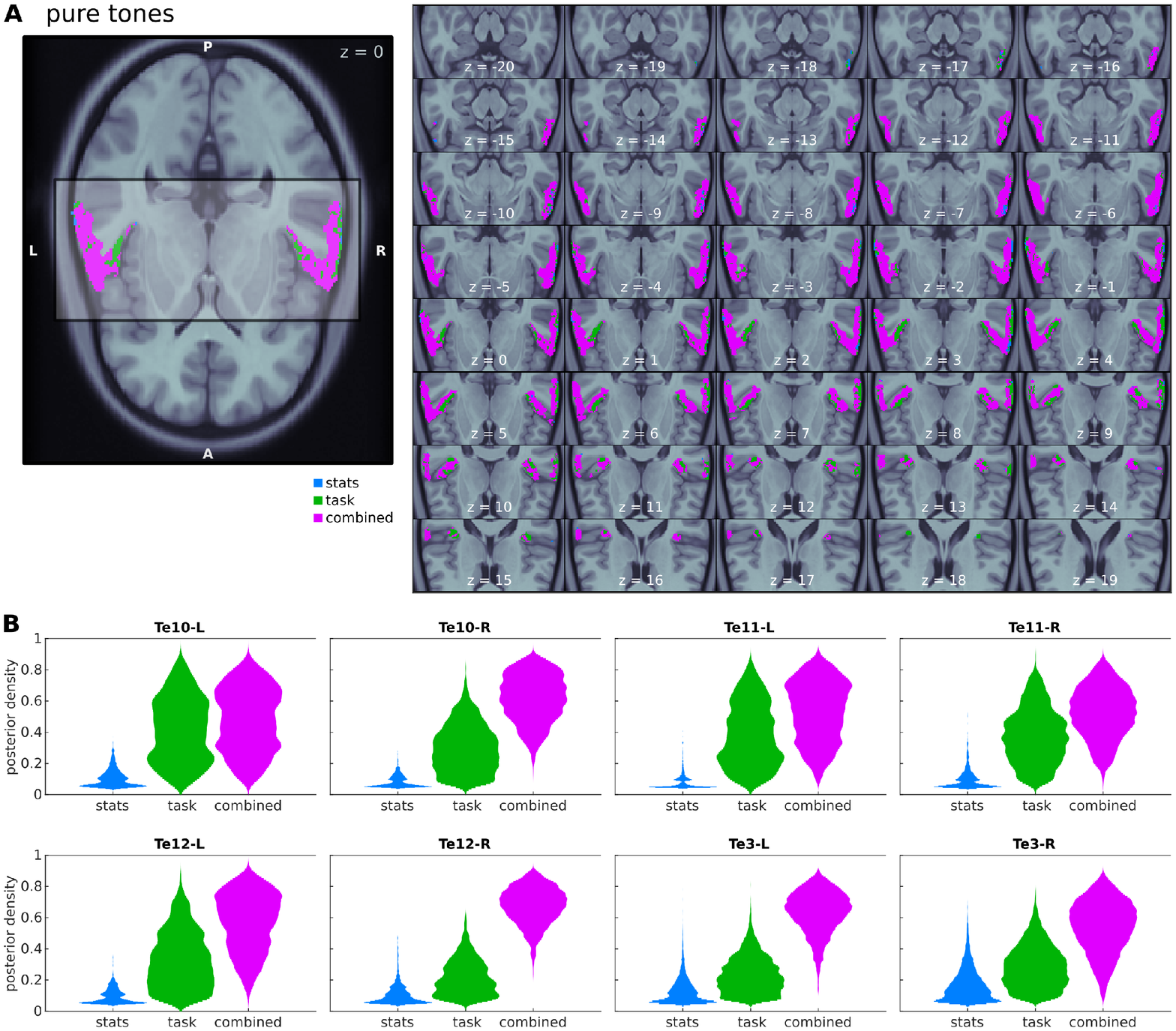}
            \caption{\textbf{Prevalence of each model in auditory cortex for pure tones.} A) Map detailing which BMC model best explains the responses to pure tones in each of the voxels of the auditory cortex. Colours indicate the BMC model with the highest posterior density at each voxel. Blue voxels are best explained by the \emph{stats-informed} BMC model, green voxels by the \emph{task-informed} BMC model, and purple voxels by the BMC \emph{combined} model. B) Distributions (kernel-density estimations) of the posterior densities of each BMC model across voxels of each of the cortical fields for the pure tone stimuli.
            \label{fig:bestmodelACpt}}
        \end{figure}

\begin{figure}[p]
            \centering
            \includegraphics[width=\textwidth]{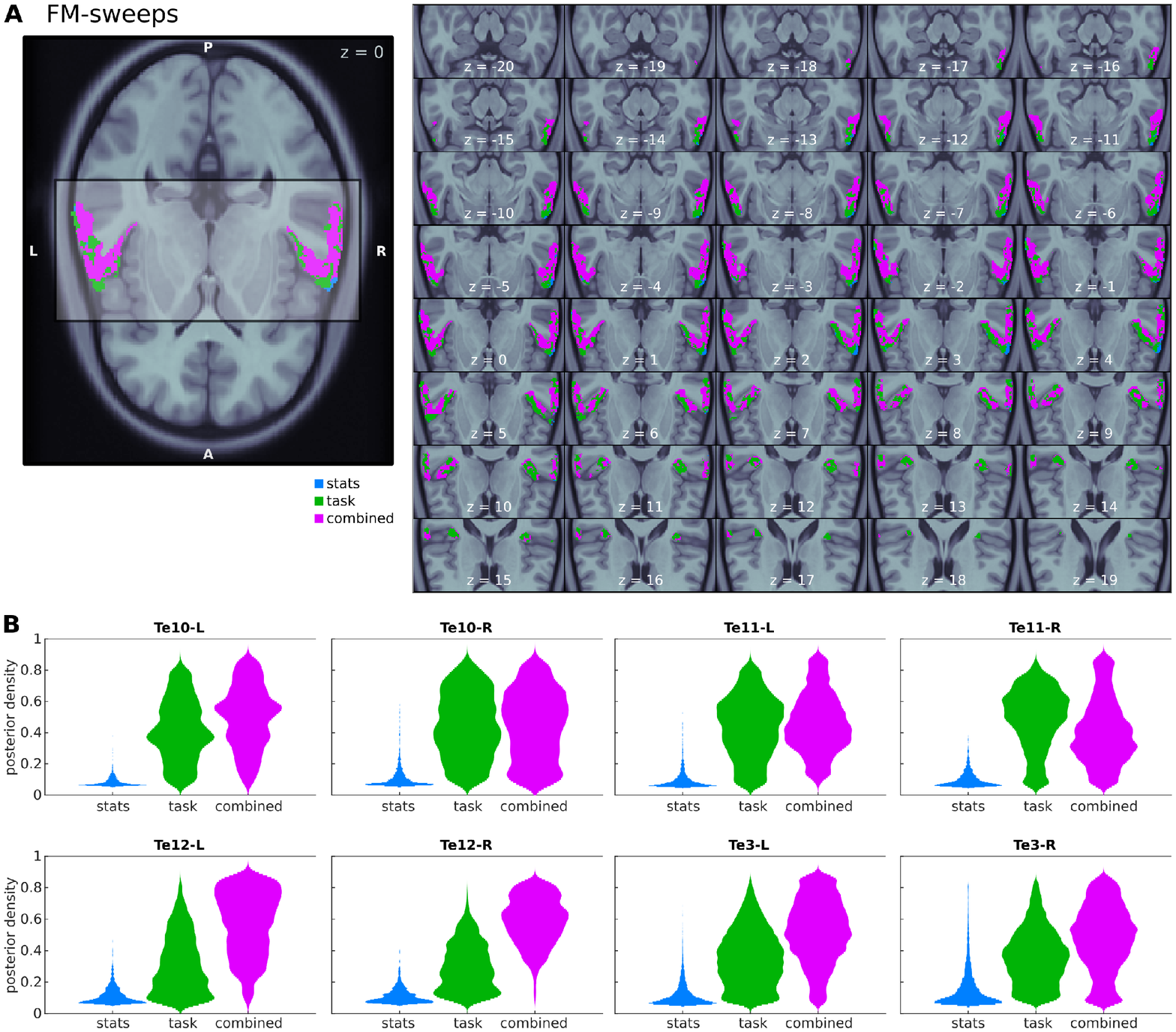}
            \caption{\textbf{Prevalence of each model in auditory cortex for FM-sweeps.}  A) Map detailing which BMC model best explains the responses to FM-sweeps in each of the voxels of the auditory cortex. Colours indicate the BMC model with the highest posterior density at each voxel. Blue voxels are best explained by the \emph{stats-informed} BMC model, green voxels by the \emph{task-informed} BMC model, and purple voxels by the \emph{combined} BMC model. B) Distributions (kernel-density estimations) of the posterior densities of each BMC model across voxels of each of the cortical fields for the FM-sweep stimuli.
            \label{fig:bestmodelACsw}}
        \end{figure}

\begin{figure}[htb!]
            \centering
            \includegraphics[width=\textwidth]{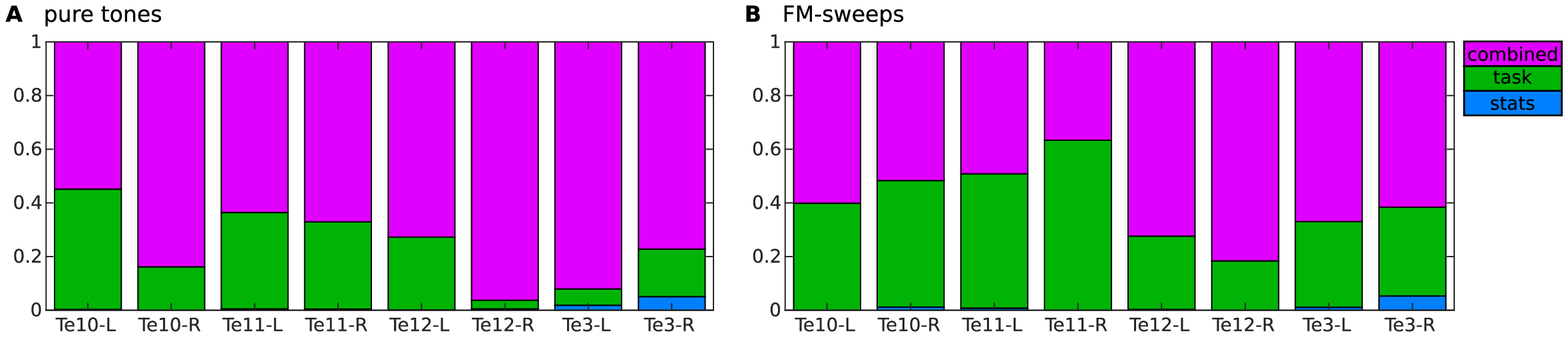}
            \caption{\textbf{Prevalence of each model in each cortical field.} Bars show the prevalence of each of the BMC models across cortical fields for the pure tone (A) and FM-sweep (B) data. Blue bars correspond to voxels that are best explained by the \emph{stats-informed} BMC model, green bars to voxels best explained by the \emph{task-informed} BMC model, and purple bars to voxels best explained by the \emph{combined} BMC model. 
            \label{fig:corticalRatios}}
        \end{figure}

        Most of auditory cortex (Te1.0, Te1.1, Te1.2, Te3, \cite{Morosan2001} was best explained by the \emph{combined} BMC model. The combined model was the best explanation of the data in more than half of the voxels across fields for pure tones (minimum of 55\% in the Te1.0-L, maximum of 96\% in Te1.2-L; Figure~\ref{fig:bestmodelACpt}) and FM-sweeps (minimum of 37\% in Te1.1-R, maximum of 82\% in Te1.2L; Figure~\ref{fig:bestmodelACsw}; exact ratios are shown in Figure~\ref{fig:corticalRatios}). The \emph{task-informed} model explained the responses of most of the remaining voxels, whilst the \emph{stats-informed} BMC model was present only minimally in two the ROIs in the pure-tone dataset (2\% and 5\% of the voxels of Te3-L and Te3-R, respectively), and four ROIs of the FM-sweep dataset (1\% of Te1.0-R, Te1.1-L and Te3-L, and 5\% of Te3-R).

        To study whether the presence of the \emph{task-informed} BMC model could also be related to the variations of the tSNR across the ROIs, we also computed the correlation between the tSNR and the posterior density of the \emph{combined} BMC model across the cerebral cortex ROIs. In the pure tone data, the posterior density was positively correlated to the tSNR in Te1.0-R, Te1.1-R, Te1.2-L and bilateral Te3 ($\rho \in [0.13, 0.45], p < 10^{-7}$), but not in the remaining ROIs ($\rho \in [-0.24, 0.07]$); in the FM-sweep data, correlations were significant in all cerebral cortex ROIs ($\rho \in [0.09, 0.70], p < 0.02$).
 
        To quantify if, as in the subcortical nuclei, the cortical organisation of the \emph{combined} and \emph{task-informed} BMC models was consistent for both stimulus families across cortical fields, we computed the correlation between the Bayes' factor $K_{combined/task}$ associated to the responses to pure tones and FM-sweeps. We found significantly positive correlations in four of the cortical fields (Te1.0-R, bilateral Te1.1, and Te3-L; $\rho \in [0.04, 0.32], p < 0.002$). However, the tSNR of the pure tone and FM-sweep datasets was also positively correlated ($\rho \in [0.31, 0.73], p < 10^{-25}$) across all cortical fields but Te1.1-R ($\rho = -0.07$), indicating that the correlations of $K_{combined/task}$ might be driven by the tSNR.

\section{Discussion}

    Previous empirical and theoretical studies on predictive coding have not considered how predictions from independent generative models are combined to compute prediction error along the processing hierarchy. Using an experimental paradigm designed to elicit two independent sets of predictions, we have shown that neural populations of the human IC, MGB and AC encode prediction error to a combination of the available predictions. Results were replicated using two families of auditory stimuli: pure tones and FM-sweeps. Our findings suggest that the auditory pathway makes use of the intricate set of corticofugal connections that transcend the linear transmission of expectations generally assumed in previous formulations of predictive coding. 
    
    \paragraph{} 
    Although predictive coding is usually formulated as a hierarchical theory comprising many processing stages \cite{Rao1999, Friston2003, Friston2009, Spratling2017, Keller2018}, previous experiments in the auditory pathway have focused on experimental setups that included a single generative model \cite{Casado2020, Malmierca2019, Parras2017, Nieto2016, Natan2015, Chen2015, Duque2015, Ayala2015b, Duque2014, Gao2014, Richardson2013, Ayala2013, Perez2012, Duque2012, Zhao2011, Antunes2011, Antunes2010, Anderson2009, Malmierca2009, Behrens2009, Ulanovsky2004, Ulanovsky2003, Heilbron2020, Tabas2021c, Tabas2020, Font2020, Cacciaglia2015, Cornella2015, Escera2014, Bendixen2012, Grimm2011, Slabu2010}. These studies did not intend to characterise the transmission of predictions from a multi-level hierarchical perspective, and so the question remained of how generative models at different stages of the hierarchy were tested by prediction error units at the lower stages.
     
    Our results indicate that BOLD responses in IC, MGB, and AC encode prediction error with respect to two generative models. This result can be interpreted in two ways. One possibility is that predictions from the available generative models are integrated into a single prediction. This could be achieved in two different ways: either by an upstream propagation of predictions that make predictions consistent across the hierarchy \cite{Spratling2017, Friston2009}, or by a (e.g., linearly weighted) integration of local and downstream propagated predictions at each stages. If predictions were transmitted downstream and integrated at the populations encoding the stats-informed model, a violation of the predictions of the integrated model could mean that the local model is wrong, that the down-propagated model is wrong, or that both models are wrong. It is unclear how the processing hierarchy could disambiguate between these scenarios to induce the necessary model updates. If predictions were transmitted upstream and integrated at the populations encoding the task-informed model, then a prediction error signal would always trigger model updates at the highest level, which captures all available information on the sensory world. But then, if prediction errors encoding violations to predictions of the stats-informed model trigger model updates at the populations encoding the task-informed model, how is the integrity of the latter preserved along time? Unlike the stats-informed model, the task-informed model cannot be reinforced by monitoring the local statistics of the activity at the immediately lower prediction error units, and information related to the task-informed model would eventually vanish from the brain. 
    
    A second possibility that is that each processing stage contains several populations of prediction error units, and that each of these populations tests the down-propagated predictions of one of the available generative models. The corticofugal bundles that directly connect the AC with the MGB, IC, and superior olivary complex \cite{Lee2011} might be responsible for the transmission of these predictions. This scenario would allow independent generative models to be maintained and updated separately and does not assume the existence of hidden states that, by definition, cannot be measured using neuroimaging techniques \cite{Friston2009}. However, much higher spatial resolution than used in the present study, possibly even single-neuron recordings, would be necessary to demonstrate that different populations at each processing stage specialise on different generative models.

    \paragraph{} 
    We found that the \emph{combined} BMC model was the best explanation for the data in AC. If we assume that the stats-informed generative model is computed at a lower stage of the processing hierarchy, this would talk in favour of the interpretation that predictions are transmitted upstream and integrated at the highest level of the processing stage. However, given the relatively high $\text{ISI} = 700$\,ms, encoding the stats-informed model might require long temporal constants that are not available at subcortical stages \cite{Steadman2018}; namely, it is likely that both, the stats- and the task-informed model are encoded in or at a higher level than AC.  
    
    \paragraph{} 
    The auditory pathway is physiologically and functionally organised in primary (or lemniscal) and secondary (or non-lemniscal) subdivisions. Primary subdivisions present narrowly tuned frequency responses and are responsible for the transmission of bottom-up information; secondary subdivisions present wider tuned frequency responses, are heavily targeted by descendent projections, and are thought to be involved in multisensory integration \cite{Hu2003}. Neural recordings from rodents suggest that prediction error may be encoded exclusively in the non-lemniscal pathway \cite{Perez2012, Duque2012, Duque2014, Ayala2015b, Malmierca2015, Nieto2016, Parras2017}. In contrast, results from human neural populations indicate that prediction error is similarly present in both, primary and secondary subdivisions \cite{Cacciaglia2015, Tabas2020, Tabas2021c, Stein2021}. 
    
    Here, we found that neural populations in the primary and secondary pathway could use different generative models to compute prediction error. This observation offers a potential explanation for why prediction error is restricted to the secondary pathways in rodents but not in humans: structures in the primary pathway do not have access to the predictions of the low-level model. Presuming that rodents make use of stats-informed models only, they would depend on the secondary structures, which have access to this model, to compute prediction error. Presuming that humans monitor the stimulation history, they might integrate the content of the stats-informed model into higher-level models (e.g., by reasoning that the tone that is repeated a lot will continue being repeated), therefore making it available to both, primary and secondary structures. However, this hypothesis is not entirely consistent with the current data: both the \emph{combined} and \emph{task-informed} models were prominent in the voxels included in our best approximation to-date of the ventral (primary) MGB in humans. We also did not observe a quantitatively larger prevalence of the \emph{combined} model in secondary areas of the auditory cortex (Te1.2, Te3). Further work is necessary to clarify why prediction error is restricted to the secondary pathway only in rodents.

    \paragraph{} 
    It might be tempting to hypothesise that the \emph{stats-informed} model reflects habituation, and that the \emph{combined} model encodes a combination of prediction error with respect to the \emph{task-informed} model plus synaptic habituation. However, the \emph{stats-informed} BMC model does not really capture habituation dynamics; rather, it assumes that the responses to the deviant will be stronger the larger the mismatch between deviant and standard $\Delta$. 
    
    If the receptive fields activated by the deviant and standard overlap significantly, habituation to the standard will result in attenuated responses to the deviant, and the attenuation would be larger the smaller the $\Delta$. However, if the deviant and standard do not share a significant segment of the receptive fields, habituation to the standard would not affect the deviant regardless of $\Delta$. In the pure tone experiment, the ERB bandwidth expected around our stimulation average frequency $\|f|\sim 1.5\,$kHz is of around $ERB \simeq 37.7$\,Hz \cite{Glasberg1990}. Therefore, although the receptive fields might overlap subtly for deviant-standard combinations at the smallest $\Delta = 45$\,Hz, habituation to the standard is unlikely to affect the responses to the deviant for $\Delta = 100$\,Hz or $\Delta = 145$\,Hz. An even stronger argument can be made for the FM-sweep data, where deviant-standard combinations included FM-sweeps with opposite FM-modulation direction.
    
    Moreover, since habituation is a passive process and an ubiquitous phenomenon in neural systems \cite{Friauf2015}, we would expect habituation to affect all voxels in the measured ROIs equally. However, the \emph{combined} BMC model is the best explanation for the data only in some segments of the ROIs, even in regions for which there was no apparent correlation between the posterior density for the \emph{combined} model and the tSNR (e.g.,~Te1.1-R, where the correlation between $K_{comb/task}$ and the tSNR was $\rho<0$, but the \emph{task} BMC model was the best explanation of the data for around a third of the voxels; Figure~\ref{fig:corticalRatios}).
    
    It could also be hypothesised that the \emph{task-informed} BMC model could be the best explanation for the data if the responses were modulated by attention-driven amplification. Tones in positions 4 and 5 are the most relevant for the task, since participants do not expect to find deviants in positions 1-3 or 7-8, and deviants in position 6 are always expected. If the responses were simply amplified by attention, the \emph{task-informed} BMC model, where responses to positions 4 and 5 are higher than to the remaining tones, would explain the data better than the \emph{stats-informed} BMC model, where all deviants elicit the same response. However, we showed in previous analyses that none of the datasets can be explained as a result of selective attention \cite{Tabas2020, Tabas2021c}. First, responses to deviants in positions 4 and 5, where participants were expected to show the same attention engagement, were significantly different and scaled by predictability \cite{Tabas2020, Tabas2021c}. Second, the magnitude of the responses to deviants in position 6 and standards in positions 7 and 8 were statistically indistinguishable, even though a previous study with a lower statistical power showed that, under passive listening, responses to deviants were always higher than to standards \cite{Cacciaglia2015}. The only explanation compatible with the data is that responses predominantly encode prediction error with respect to the \emph{task-informed} generative model.

    \paragraph{} 
    Our results show a generally higher prevalence of the \emph{combined} BMC model in pure tones than in FM-sweeps. However, the distribution of $K$ to the contrast between the \emph{combined} and \emph{stats-informed} models is significantly correlated between pure tones and FM-sweeps in most of the ROIs included in the study. These effects are likely driven by the tSNR heterogeneity of the data, given the high correlations we found between the posterior density of the \emph{combined} BMC model and the tSNR.
    
    Voxels encoding prediction error with respect to the \emph{combined} BMC model are more elusive in the FM-sweep data because modulation rate is not enough to characterise FM, therefore decreasing the fitting power of the model we used to encode $\Delta$ in FM-sweeps. Direction and rate are indeed typically studied as independent features in the literature \cite{Lui2003, Geis2013, Issa2016, Hsieh2012, Altmann2014}, and single neurons in the auditory pathway are usually either selective to FM-direction or to FM-rate. Therefore, FM might be encoded in a two-dimensional feature space in the brain. A possible solution to this problem could have been to incorporate a contribution of FM-direction to $\Delta$ as an extra parameter in the BMC models used to analyse the FM-sweep data. However, adding yet-another extra regressor would have increased the dependence of the log-evidence of the \emph{combined} model on the tSNR even further.

    \paragraph{} 
    
    Predictive coding has gone a long way since it was first explicitly theorised in the 90s \cite{Mumford1992}. Here we have taking a step forward by investigating, for the first time, how predictive processing generalises to a scenario where the sensory world is predicted by two concurrent independent generative models. Understanding predictive coding from a multi-level perspective is crucial to understand how natural sensory processing occurs. For instance, predictive speech processing requires integrating predictions stemming from contextual information, semantics, grammatical constraints, phonetic rules, and the frequency register of the speaker \cite{Choi2021, Heilbron2020, Kuperberg2016}. To extract meaningful messages from noisy and ambiguous speech signals, the human brain should be able to exploit information from all those independent models. Our current findings suggest that, at least at low-level stages of the processing hierarchy, prediction error units exploit this information in a  synergistic fashion. The auditory pathway might exploit the corticofugal lines directly connecting the AC with the MGB, IC and superior olivary nucleus for the direct transmission of predictions, bypassing the linear hierarchy often assumed in the literature \cite{Keller2018}. This complex intricate descending system of connections might be responsible for our fine-tuned capacity to sense that that we are able to predict.


\bibliographystyle{ieeetr}
\bibliography{bib}

\end{document}